\documentclass[]{tMOP2e}
\usepackage{epstopdf}
\usepackage{color}
\begin{document}
\doi{10.1080/0950034YYxxxxxxxx}
\issn{1362-3044}
\issnp{0950-0340} \jvol{00} \jnum{00} \jyear{2012} \jmonth{1 September}
\markboth{Taylor \& Francis and I.T. Consultant}{Journal of Modern Optics}

\title{Coupling a small torsional oscillator to large optical angular momentum}

\author{H. Shi and M. Bhattacharya$^{\ast}$\thanks{$^\ast$Corresponding author. Email: mxbsps@ rit.edu
\vspace{6pt}}\\\vspace{6pt}  {\em{School of Physics and Astronomy, Rochester Institute of Technology, 84 Lomb Memorial
Drive, Rochester, NY 14623}}
\\\vspace{6pt}\received{September 28th, 2012}}

\maketitle

\begin{abstract}
We propose a new configuration for realizing torsional optomechanics: an optically trapped windmill-shaped dielectric
interacting with Laguerre-Gaussian cavity modes containing both angular and radial nodes. In contrast to existing schemes,
our method can couple mechanical oscillators smaller than the optical beam waist to the in-principle unlimited orbital
angular momentum that can be carried by a single photon, and thus generate substantial optomechanical interactions.
Combining the advantages of small mass, large coupling, and low clamping losses, our work conceptually opens the way
for the observation of quantum effects in torsional optomechanics.
\bigskip

\bigskip
\end{abstract}
\vspace{6pt}

\section{Introduction}
Recent advances in experimental capabilities have granted physicists access to the quantum mechanical behavior of macroscopic
harmonic oscillators \cite{Kippenberg2008,Marquardt2009,Schnabel2011}. The basic paradigm underlying many of the
spectacular demonstrations along this line of research stems from the interaction between a linearly vibrating mechanical
object and a high finesse optical mode. Exploiting this optomechanical coupling, experimentalists have been able to measure
oscillator displacement with an imprecision below the quantum limit \cite{Teufel2009, Braginsky}, engineer the quantum ground state of the oscillator
\cite{Painter2011}, and discern the effects of the quantum back-action of radiation pressure \cite{Kurn2012}. A large number of
theoretical proposals, targeting squeezing \cite{Nunnenkamp2010}, quantum superposition \cite{Bouwmeester2008}, entanglement \cite{Vitali2007}, information processing \cite{Tombesi2003, Rabl2012}, and exploration of the quantum-classical boundary \cite{Cirac2011} have also been put forward. In general, quantum effects are easier to detect in optomechanical systems if optical finesse, mechanical quality, mechanical frequency, and optomechanical coupling
are high, and oscillator mass and ambient temperature are low \cite{Kippenberg2008}.

In the last few years, the extension of optomechanics from linear to torsional oscillations has been considered
\cite{Tittonen1999, Bhattacharya20072, Wang2008, Isart2010}. This effort has been motivated by the possibility of producing
rotation sensors \cite{Wang2008}, by interest in the role of photonic orbital angular momentum in manipulating matter
\cite{Photonsbook}, and by the desire to optomechanically address particles with more than one centre-of-mass degree of
mechanical freedom \cite{Isart2010}. Experimentally, torsional optomechanical oscillators have been presented, which however
relied on linear optical momentum, which
is difficult to scale up in practice \cite{Tittonen1999, Wang2008}. These oscillators also therefore did not exploit the large orbital angular
momentum that can be carried by a single photon. On the theoretical front, a Fabry-P\'{e}rot design has been suggested, using spiral
phase plates as end mirrors, and based on the exchange of orbital angular momentum between the intracavity beam and one of the
end plates mounted as a torsional oscillator \cite{Bhattacharya20072}. However, it is presently difficult to produce such
cavities with high finesse in the laboratory. Also, losses due to clamping introduce undesirable mechanical dissipation to
the system.

A subsequent pioneering proposal brought torsional optomechanics within the realm of present experimental capabilities by
theoretically considering a sub-wavelength-sized dielectric inside a standard high finesse cavity made of spherical mirrors \cite{Isart2010}.
Optical trapping of the dielectric was proposed to avoid clamping losses, while the rotational degree of freedom was addressed
by Laguerre-Gaussian optical modes $LG_{l,p}$ carrying orbital angular momentum $l \neq 0$ but with a radial index $p=0$.
Couplings both linear (for cooling) and quadratic (for trapping) in the angular displacement were shown to be theoretically
achievable. We show below that such couplings decrease rapidly with increasing $l$, requiring the impracticable use of high
optical power for observing optorotational effects.

In this article, we suggest instead an optomechanical configuration that couples the torsional vibrations of a small, optically
trapped, windmill-shaped dielectric to Laguerre-Gaussian $LG_{l,p}$ cavity modes carrying angular $(l \neq 0)$ as well as radial
$(p\neq 0)$ nodes. We show that such a configuration produces, for a dielectric equal to or smaller in size than the optical beam waist,
an optomechanical interaction which increases with $l$, provided $p$ is chosen appropriately. The advantages of small mass,
large coupling, and low clamping losses should facilitate the observation of quantum effects in the proposed system. We note that Laguerre-Gaussian beams with radial nodes have been used to rotate nanowires recently in an optical tweezer experiment \cite{Wu2012}, and also
have been analyzed for tweezing efficiency \cite{Wang2012}.

\section{Physical system}
\begin{figure}
\begin{center}
\includegraphics[width=3in]{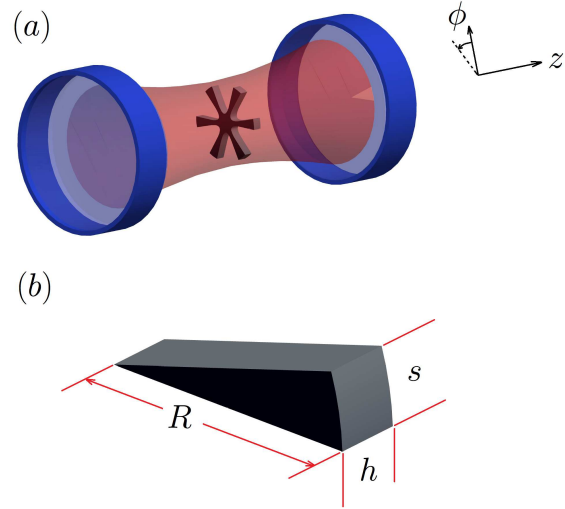}
\caption{\label{fig:config}The physical configuration we suggest for torsional optomechanics. (a) A windmill shaped object optically trapped
in a high finesse optical cavity. The object can undergo torsional oscillations along the angle labelled as $\phi$. The rod is
trapped near the centre of the cavity. For viewing clarity, the details of mode profile of the Laguerre-Gaussian cavity mode have not been shown here. (b) One of the wedges with its dimensions labelled.}
\end{center}
\end{figure}

The physical system we propose is shown in Fig. \ref{fig:config}. A windmill shaped object is placed inside a high finesse optical
cavity and interacts with Laguerre-Gaussian modes supported by the cavity. The windmill is composed of $l$ spokes, each of which
includes two wedges with radius $R$, arc length $s$, and thickness $h$. The moment of inertia of the dielectric about its geometric
centre is $I=lmR^2$. Such windmill rotors have been micromachined and rotated optically in the laboratory \cite{Galajda2002}.
We note that the configuration suggested in Ref. \cite{Isart2010} uses only one spoke; our proposal serves to increase the net
optomechanical coupling.

We assume that the windmill is trapped in harmonic potentials axially at $z_0\sim0$, radially
at $r_0\sim0$, and azimuthally at $\phi_{0} \sim 0$, where the trapping frequency in the $\phi$ direction is $\omega_\phi$. The
trapping can be realized with either optical tweezers \cite{Li2011, Law2012, Chang2009, Singh2010, Barker2010, Schulze2010, Feng2011} or a self-trapping scheme using modes of the optical cavity
\cite{Isart2010}. The specifics of the trapping configuration are not important for our analysis, therefore we do not discuss them in detail in
the present article. The optomechanical coupling results from the interaction between the windmill and an optical field produced
by the superposition of two degenerate but counter-rotating cavity modes $LG_{l,p}$ and $LG_{-l,p}$, with a beam profile given by
\cite{SiegmanBook}
\begin{eqnarray}
\label{eq:LGmode}
|\psi_{l,p}({\bf r})|^{2}&=&A_{l,p}\frac{w_{0}^{2}}{w^{2}(z)}\left[\frac{r\sqrt{2}}{w(z)}\right]^{2|l|}
\exp\left(-\frac{2r^{2}}{w^{2}(z)}\right)L_{p}^{|l|}\left[\frac{2r^{2}}{w^{2}(z)}\right]^{2}\cos^{2}kz\cos^{2}l\left(\phi-\phi'\right),
\end{eqnarray}
In Eq.(~\ref{eq:LGmode}), $\psi_{l,p}({\bf r})$ is the optical mode function, $A_{l,p}=2p!/[(1+\delta_{l,0})(|l|+p)!]$ is a normalization
factor, $w(z)=w_{0}\sqrt{1+(z/z_{R})^{2}}$ is the beam diameter, where $w_{0}$ is the beam waist, $z_{R}=\pi w_{0}^{2}/\lambda$ is the
Rayleigh range, and $\lambda$ is the wavelength of the optical radiation. Also, $L_{p}^{|l|}$ is an associated Laguerre polynomial, and $\phi'$
is the relative phase between the two modes. We note that the validity of Eq. (\ref{eq:LGmode}) requires the Rayleigh range $z_{R}$ to be
larger than the cavity length $L$.

Both the radial and the axial motions of the dielectric are completely decoupled from the windmill's azimuthal motion, and thus may be
ignored. We will consider only the torsional interaction between the dielectric and the Laguerre-Gaussian beams. This interaction can be
arranged to be linear in the angular displacement by choosing $\phi' = \pi/4l \sim 15\,^{\circ}$ \cite{Isart2010}, and can be modeled by the Hamiltonian
\cite{Bhattacharya2007,Kippenberg2008}
\begin{equation}
\label{eq:HamSensor}
H=\hbar \omega_{c}(\phi_{0}) a^{\dagger}a+ \hbar \omega_\phi b^{\dagger}b+ \hbar g_{l,p} a^{\dagger}a (b+b^{\dagger}),
\end{equation}
where $a(a^{\dagger})$ and $b(b^{\dagger})$ are the annihilation (creation) operators of the $\psi_{l,p}({\bf r})$ mode and the mechanical
mode obeying the bosonic commutation rules $[a,a^{\dagger}]=1$ and $[b,b^{\dagger}]=1$, respectively. The coupling constant $g_{l,p}$
in Eq. (\ref{eq:HamSensor}) can be calculated from the $\phi$-dependent cavity resonant frequency $\omega_c(\phi)$, whose departure
from the equilibrium value $\omega_c(\phi_0)$ due to the presence of the dielectric can be estimated using perturbation theory, since the
dimensions of the dielectric are smaller than those of the beam $(s < \lambda, h < D, R \lesssim w_{0}/2)$ \cite{Johnson2002},
\begin{equation}
\label{eq:FreqMode}
\frac{\omega_{c}(\phi)}{ \omega_{c}(\phi_{0})}\simeq1-\frac{\int_{V} (\epsilon-1)|\psi_{l,p}({\bf r})|^{2}d {\bf r}}
{2\int_{V'} |\psi_{l,p}({\bf r})|^{2}d {\bf r}},
\end{equation}
where $\epsilon$ and $V$ are the dielectric constant and volume, respectively, of the windmill, and $V'$ is the volume of the whole cavity. The
optomechanical coupling in Eq.~(\ref{eq:HamSensor}) can then be found as
$g_{l,p}=\sqrt{\hbar/I \omega_\phi}\partial \omega_{c}(\phi)/\partial \phi |_{\phi_{0}}$ from Eq.~(\ref{eq:FreqMode}).

\section{Laguerre-Gaussian modes without radial nodes}
\begin{figure}[t]
\begin{center}
\includegraphics[width=3in]{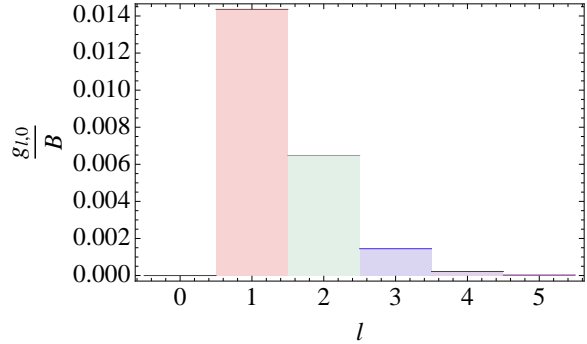}
\caption{(Color online) Plot of the optomechanical coupling [Eq.~(\ref{eq:GL0})], between the dielectric windmill of Fig.~\ref{fig:config} and the optical mode of Eq.~(\ref{eq:LGmode}) for $R/w_{0}=1/2, p=0$, and various $l$.}
\label{fig:nop}
\end{center}
\end{figure}

\begin{figure}
\begin{center}
\includegraphics[width=3in]{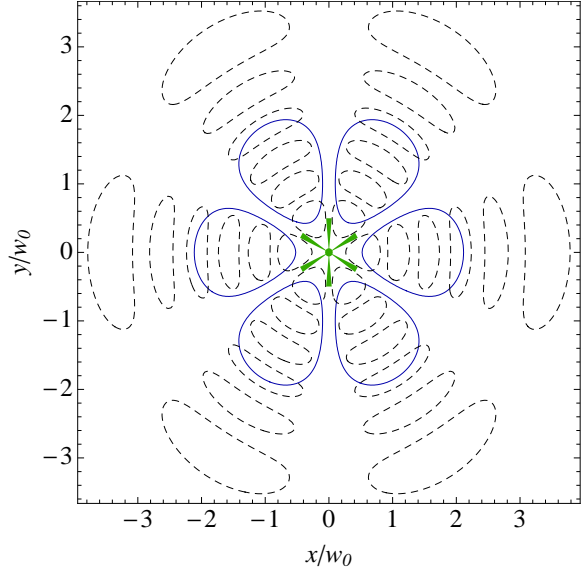}
\caption{(Color online) A cross-section of the cavity of Fig.~\ref{fig:config} at $z=0$, showing the windmill dielectric
and the optical modes (with a beam waist of $w_0$) it interacts with. The intensity maxima drawn with solid lines
correspond to Eq.~(\ref{eq:LGmode}) with $l=3,p=0$, while the lobes drawn using dashed lines correspond to
Eq.~(\ref{eq:LGmode}) with $l=3,p=5$. For clarity of viewing, $\phi'=0$ has been used in both cases.}
\label{fig:profile}
\end{center}
\end{figure}

In this section, we consider the optomechanical coupling $g_{l,p}$ in detail. We first assume $l \neq 0, p=0$, which corresponds to a beam profile without radial nodes.
Generalizing the treatment of Ref.~\cite{Isart2010} to arbitrary $l$, we find
\begin{equation}
\label{eq:GL0}
\frac{g_{l,0}}{B}=\frac{\Gamma\left[|l|+1\right]-\Gamma\left[|l|+1,2\left(\frac{R}{w_{0}}\right)^{2}\right]}{|l|^{-1}(|l|-1)!(1+\delta_{l,0})},
\end{equation}
where $\Gamma[|l|+1]$ is the complete and $\Gamma[|l|+1,2R^{2}/w_{0}^{2}]$ the incomplete Gamma function, and $B=(\epsilon-1)(wt/\pi RD)\omega_{c}\left(\phi_{0}\right)\sqrt{\hbar/mR^{2}\omega_\phi}$ is a constant independent of $l$. Choosing
$R/w_{0}=1/2$ as in Ref.~\cite{Isart2010}, we see from Fig.~\ref{fig:nop} that $g_{l,0}/B$ decreases dramatically with $l$. This
behavior can be understood from the simple fact that the single radial maximum of the $|\psi_{l,p=0}({\bf r})|^{2}$ mode profile
near $z\sim 0$ lies at $r_{\mathrm{max}}^{|l|}\simeq w_{0}\sqrt{|l|/2}$ \cite{Dholakia2001}. Thus, for higher values of $l$ the
radiation is too far away from the beam centre $(r_{\mathrm{max}}^{|l|} \gg R)$ to couple to any dielectric equal to or smaller
than the beam waist $(R \lesssim w_{0}/2)$. Such a dielectric therefore experiences mostly the dark core of the optical mode. This
situation has been depicted in Fig.~\ref{fig:profile}, which shows a cross-section of the cavity at $z=0$ in the $x-y$
plane. Clearly, the windmill overlaps very little with the six intensity maxima for the $l=3, p=0$ beam shown using solid
curves.

One way of coupling to higher $l$ modes is, of course, to increase proportionately the radius $R$ of the windmill. However, the
corresponding increase in mass would make the observation of quantum effects, as well as optical trapping, more challenging;
the increase in surface area would
lead to larger mechanical damping rates due to collisions with background gas molecules; and the increase in size would lead to more
optical scattering due to diffraction, reducing the finesse of the cavity. For these reasons, we consider it important to find a
technique for strengthening the optomechanical coupling without using dielectrics with diameter larger than the optical beam
waist $w_{0}$(a parameter which can be held constant as $l$ is increased). Below, we suggest such a method.
\section{Laguerre-Gaussian modes with radial nodes}
\begin{figure}
\begin{center}
\includegraphics[width=3in]{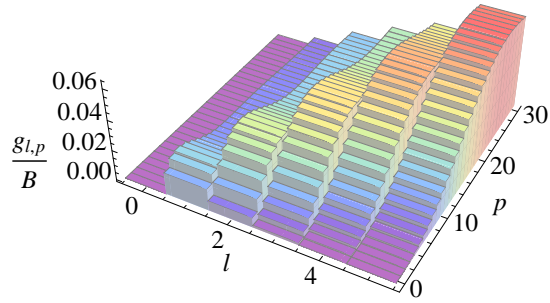}
\caption{(Color online) Plot of the optomechanical coupling $g_{l,p}/B$, between the dielectric windmill of Fig.~\ref{fig:config}
and the optical mode of Eq.~(\ref{eq:LGmode}) for $R/w_{0}=1/2$, and various $l$ and $p$.}
\label{fig:withp}
\end{center}
\end{figure}

Consider now a situation in which $l \neq 0, p \neq 0$. Laguerre-Gaussian cavity modes with $p$ up to 12 and $l$ up to 28 have been produced
experimentally \cite{Ueda2010}. In this case, $g_{l,p}/B$ could be calculated only
numerically for arbitrary $l$ and $p$, and is displayed in Fig.~\ref{fig:withp} \footnote{We note that $g_{l,p}/B$ can be calculated
analytically for any given $l$, and arbitrary $p$.}. This figure shows two clear trends. First for
every $l$, there is a value of $p$ at which the coupling reaches a maximum. Second, and in sharp contrast to
Fig.~\ref{fig:nop}, the magnitude of this maximum increases with $l$. These characteristics can be broadly understood from
Fig.~\ref{fig:profile}, which displays the thirty six intensity maxima of the $l=3, p=5$ beam as dashed curves. Clearly, for
larger $p$, the six innermost lobes of the beam overlap better with the dielectric, leading to an increase in the optomechanical
coupling. However, the appearance of thirty additional lobes results in loss of intensity from the innermost maxima, thus weakening
the coupling with the dielectric. The competition between these two opposing mechanisms yields, for a given $l$, an optimum
$g_{l,p}$ at a particular value of $p$. For $l=3$, for example, the peak in the coupling occurs at $p=11$, as can be seen from
Fig.~\ref{fig:withp}. For higher values of $l$, the maximum coupling is stronger, since the slope of the mode
profile $|\psi_{l,p}({\bf r})|^{2}$ near the angular equilibrium position is sharper along $\phi$ in this case.

Assuming that the windmill is machined from fused silica, and using parameters from relevant experiments \cite{Novotny2012, Raizen2011}
and experimental proposals \cite{Isart2010} ($D=0.5$ mm, $\lambda=1064$ nm, $w_0=20$ $\mu$m, $l=3, p=11$,
$\epsilon=2.1$, $m=10^{-16}$ kg, $R=10~\mu$m, $s=h=200$ nm, $\omega_{\phi}=50$ kHz), we find $g_{3,11} \sim 200$ Hz.
This is much larger than the coupling for a $p=0$ beam, $g_{3,0}\sim 5$ Hz. Clearly, the presence of radial nodes increases
the coupling dramatically. Further, reducing the radius to $R=8~\mu$m we find a coupling of $100$ Hz by using $l=3$ and $p=15$.
The method we have suggested can therefore be used in principle for dielectrics of the size of the beam waist or smaller. In
practice, the waist of a diffraction-limited optical beam is of the order of a few wavelengths, i.e. a few microns. For comparison,
optomechanical experiments have involved dielectric particles ranging in size from tens of nanometers \cite{Novotny2012} to a
few microns \cite{Raizen2011}. With currently available $LG_{l,p}$ modes, our technique should be able to address particles at
the upper limit of such length-scales.

Before concluding this section we make two observations. Our first comment relates to our choice of a windmill shape for the
dielectric. If the fabrication of the windmill proves to be difficult for high $l$ (i.e. large number of spokes), it can be
substituted by a single rod, as in Ref.\cite{Isart2010}. We have found that in this case the presence of radial nodes in the
optical mode yields an optomechanical coupling whose behavior with $l$ is qualitatively the same as for the windmill, although
its magnitude is reduced somewhat, as may be expected. Second, we note that optomechanical couplings quadratic in $\phi$, which
increase with $l$, are also achievable using our method, by setting $\phi'=0$ in Eq. (\ref{eq:LGmode}). The extension of the
self-trapping scheme of Ref. \cite{Isart2010} to beams containing radial nodes would in fact be useful to optically confine
torsionally oscillating dielectrics such as presented in this article.
\section{Decoherence}
\begin{figure}
\begin{center}
\includegraphics[width=3in]{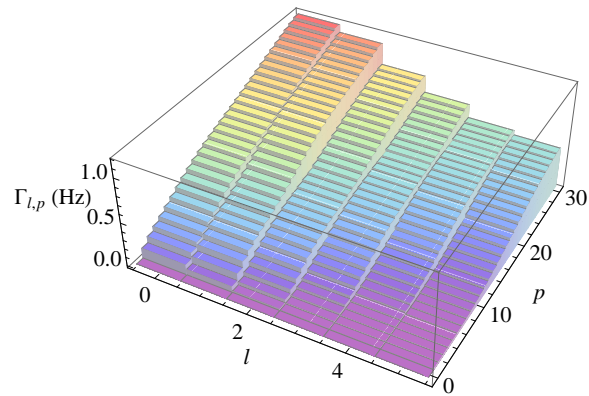}
\caption{(Color online)Plot of the decoherence rate $\Gamma_{l,p}$, due to photon scattering from the trapping beam by the
dielectric windmill of Fig.~\ref{fig:config} from the optical mode of Eq.~(\ref{eq:LGmode}) for $R/w_{0}=1/2$, and various
$l$ and $p$. The power in the trapping beam incident on the cavity is 0.1mW \cite{Pflanzer2012, Lechner2013}.}
\label{fig:dec}
\end{center}
\end{figure}

From the perspective of quantum applications, it is relevant to consider the decoherence of the dielectric due to mechanisms
such as photon scattering. The decoherence rate of the dielectric due to the scattering of photons out of the cavity can be
calculated using a master equation approach \cite{Pflanzer2012}. In Ref. \cite{Pflanzer2012}, the decoherence rate
$\Gamma^{\mathrm{cav}}$ was calculated for a dielectric sphere with radius up to twice the optical wavelength. We have extended
the calculation to a dielectric sphere with radius up to ten times the optical wavelength. For a sphere of this size, geometrical
scattering predominates. In order to adapt the decoherence calculation for the sphere to our geometry, we need to account for the
fact that our system is composed of a dielectric windmill, which has a much smaller cross sectional area than a sphere of comparable
radius. Also, our proposal involves Laguerre-Gaussian beams, which have several radial and azimuthal intensity minima, as compared
to a Gaussian beam considered in the case of the sphere. These factors are expected to reduce the photon scattering rate as compared
to dielectric spheres in Gaussian beams, and can be accounted for by multiplying $\Gamma^{\mathrm{cav}}$ by the following
geometric ratio
\begin{equation}
\zeta_{l,p}=\frac{\int_{A_{w}}|\psi_{l,p}({\bf r})|^{2}da}{\int|\psi_{l,p}({\bf r})|^{2}da},
\end{equation}
where $A_w$ is the geometric cross sectional area of the windmill. Multiplying the decoherence rate for the sphere with this
dimensionless geometric factor, we obtain a decoherence rate
\begin{equation}
\label{eq:Glp}
\Gamma_{l,p}=\zeta_{l,p}\Gamma^{\mathrm{cav}}.
\end{equation}
This decoherence rate is shown in Fig.~\ref{fig:dec}, for the same ranges of $l$ and $p$ as in Fig.~\ref{fig:withp}.
We find that typically $\Gamma_{l,p} \simeq 1$Hz, much smaller than the coherent coupling rate of $g_{l,p} \simeq 100$Hz
demonstrated above. We note here that the optomechanical coupling depends on the relative volumes of the dielectric and
the optical mode [Eq.~(\ref{eq:FreqMode})], while the photon scattering rate is governed by the ratio of their transverse
areas [Eq.~(\ref{eq:Glp})] Fig.~\ref{fig:withp}. The key point of course is that the rate of decoherence is small compared to
the coherent coupling in the regime of interest, making our proposal feasible.
\section{Conclusion}
In this article we have suggested a new torsional optomechanics configuration using the interaction between an optically trapped
``windmill"-shaped dielectric and a superposition of two counter rotating Laguerre-Gaussian modes. We have demonstrated that the
in-principle unlimited angular momentum carried by a photon can be harnessed to dielectrics smaller than the beam waist if the
number of radial nodes in the optical mode is chosen appropriately. We have shown that this scheme can provide sizable
optomechanical interactions using realistic experimental parameters. Featuring small mass, large coupling, and low clamping losses,
our proposal paves the way for exploring quantum effects in torsional optomechanics.
\section*{Acknowledgements}
We thank Dr. S. Preble and Dr. E. Hach III for useful discussions and the Research Corporation for Science Advancement
for support.
\bibliographystyle{tMOP}
\bibliography{JMOLGLP}
\vspace{36pt}
\markboth{Taylor \& Francis and I.T. Consultant}{Journal of Modern Optics}

\label{lastpage}

\end{document}